# A solution approach for the anonymous sharing of sensitive supply chain traceability data

*Rob Glew (rg522@cam.ac.uk)*
*Institute for Manufacturing, University of Cambridge*

*Ralph Tröger (ralph.troeger@gs1.de)*
*GS1 Germany*

*Sebastian Schmitter (sebastian.schmittner@eecc.de)*
*European EPC Competence Center (EECC)*

## Abstract

Supply chain traceability systems have become central in many industries, however, traceability data is often commercially sensitive, and firms seek to keep it confidential to protect their competitive advantage. This is at odds with calls for greater transparency in supply chains for sustainability and efficiency. In this research paper, we develop a solution to this conflict. We demonstrate that sanitised data can be used to confirm the authenticity of products and anonymously verify the rights of an actor to access detailed traceability data. Following a design science methodology, we develop a prototype protocol and verify its performance with traceability data based on the global standard EPCIS.

## Introduction

Product quality and safety in food supply chains is verified by traceability data. In the event of a recall, this data can reduce the time and cost for all affected products to be removed from sale (Aung & Chang, 2014). Traceability data is recorded and shared by all actors in the supply chain using a range of digital technologies. However, as the complexity of supply networks grows (as in fashion or automotive, for example), it is not possible for a firm to identify every actor that may have a right to safety or quality critical product data (Kumar et al. 2017). Sharing this data publicly risks revealing commercially sensitive information about production processes or business relationships (Massimino et al. 2018).

The current state-of-the-art falls short, both ignoring the problem of confidentiality and calling for full chain transparency to avoid it (Massimino et al. 2018). Instead, in this research paper we aim to balance the requirements for privacy and confidentiality with the practical need to achieve full supply chain traceability, support product recalls, protect consumers, and improve operational efficiency. We answer the question: *How can traceability data be shared with a priori unknown parties if the request is justified, whilst respecting data ownership and maintaining confidentiality of sensitive information, including the data owner's identity?*

To do so, we adopt a design science methodology, through which we develop a prototype software protocol based on seven functional requirements. The prototype is tested with simulated traceability data for a circular electronic products supply chain. Whilst further work is required to implement the proposed solution in an empirical setting and test its performance, with this work we demonstrate a concrete solution to anonymous, yet practical, supply chain traceability.

## Background

Supply chain traceability systems (SCTSs) enable a 'tracking and tracing' capability within a supply network (Olsen & Borit, 2013). For instance, SCTSs are commonly deployed in the food supply chain, where they provide assurance over the quality of products and improve recall processes (Aung & Chang, 2014). However, as described by Tröger et al. (2018), sharing of data in SCTSs is often limited by concerns regarding the confidentiality and commercial sensitivity of product and process data.

Traceability data is required for critical business functions, such as identifying the origin of goods and confirming their safety. This protects manufacturers and retailers from reputational damage (Aung & Chang, 2014). In some cases, traceability data must also be shared beyond the supply chain, such as for certification of sustainability in the timber industry (Appelhanz et al. 2016).

Data access is not usually a barrier in simple linear supply chains, as all the actors can be identified a priori and one firm usually dominates (such as a retailer, as in Kassahun et al. 2016). However, in more complex supply networks SCTSs are decentralised and many actors in the network are not known to each other (Lu et al. 2019). In these systems, a firm may receive requests for data from other firms who claim to handle the product and therefore have a right to part of its traceability history. Table 1 compares centralised and decentralised SCTSs. Although decentralised traceability systems have been proposed in the literature, the unwillingness of firms to share operational data with unknown parties has proved a barrier to their adoption (Biswas et al., 2017; Tröger et al., 2018).

*Table 1 – Visibility event data sharing choreographies (developed from Tröger et al. 2018)*

| Network Topology | Characteristics | Discovery Methods | Limitations | Existing Research |
|---|---|---|---|---|
| Centralised | Visibility event data is pushed to a central repository<br>Data access/exchange via a query interface | Owner of centralised repository handles discovery, data owners may set access conditions | Requires trust in centralised system<br>Not suitable in complex, emergent networks | Dominant solution explored in the extant literature<br>e.g. Kassahun et al. (2014, 2016) |
| Decentralised & Replicated | Visibility event data is placed on a ledger and all supply chain parties keep local copies of all data | Full transparency<br>Public/Private key approach | Sensitive data is replicated at nodes in the chain, may be exposed to attack even if sanitised | Interest from a technical perspective, BCT commonly applied.<br>e.g. Tian (2017) |
| Decentralised & Non-Replicated | Visibility event data is kept in the local systems of the data owner, not replicated across all nodes. | CNA (GS1, 2017) | Data distributed among an unknown number of unconnected parties | Limited consideration in academic literature, no attempts to address discovery. |

*Existing Approaches to Anonymous Data Sharing*

Solutions to anonymous data sharing and confidentiality have received inadequate attention in the fields of supply chain management and information systems generally (Massimino et al. 2018). Although some authors have proposed decentralised supply chain traceability systems, as far as we are aware none have developed a solution to anonymous traceability data discovery.

In a partial solution, Pardal et al. (2012) proposed a so-called 'Chain Navigation Algorithm (CNA)'. The CNA would automatically query its way backwards in the supply network, being provided with visibility event data that had been 'sanitised' to remove any sensitive data. However, the CNA only operates backwards in the supply network and does not support data sharing, only verifying that an unbroken chain of ownership exists.

Tian (2017) developed a concept for a system based on distributed ledger technology (DLT) which replicates visibility data at every node in the supply network. The author proposed that rules be set to govern access, but this solution would not be feasible in a supply network, because read rights to the data on the DLT cannot be controlled. Another DLT solution was proposed by Biswas et al. (2017) for the wine supply chain. In this case, the confidentiality issue was addressed by the use of a public/private key system. Some data in the decentralised ledger was encrypted to avoid sharing sensitive information, however this solution is again not applicable in complex supply networks where parties are not known. DLT systems also face limitations due to their complexity and the cost of on-chain data storage (Do & Ng, 2017).

In summary, three solution approaches exist at present. First, many authors call for transparency to support traceability and therefore avoid any problems with access rights. Second, the chain navigation approach theoretically allows a chain of custody to be confirmed but has never been executed in practice. Finally, DLT-based solutions are used to share data or proof integrity, in some cases even with some degree of privacy, but implementing access restrictions in DLT based systems is typically difficult. To overcome the limitations of these approaches, a system is required to anonymously verify access rights and share data, whilst also confirming a complete chain of custody exists. This research paper sets out such a solution.

## Method

This research paper employs the design science in information systems methodology (Hevner and Chatterjee, 2010). Design science is appropriate to answering "how" questions, by constructing and evaluating information system prototypes. We develop a prototype solution to a clearly defined problem and evaluate it against a set of functional requirements using simulated traceability data.

We first reviewed the academic and practitioner literature to clearly define the problem and identify any existing solutions or partial solutions. We then developed a set of function requirements guided by work from the GS1 Discover Service Working Group. The prototype solution was then developed in Python. Each module was individually tested before being integrated into the completed prototype. Finally, test data was used to ensure the prototype solution met the functional requirements.

The system developed for this research project is described below. Further research is required to develop and test the prototype system in an empirical setting. This work is underway.

## Solution Functional Requirements

We identified the functional requirements based on shortcomings of existing approaches (e.g. Tröger et al. 2018), industry needs (e.g. Guenther & Woerner 2019), and business requirements as gathered in the course of related standardisation work such as the GS1 Discovery Service Working Group (see GS1, 2021). Any solution must be able to achieve a high level of market acceptance and be both scalable and interoperable. Table 2 summarises the functional requirements.

*Table 2 – Functional requirements for anonymous supply chain traceability data sharing*

| # | Requirement | Explanation and Rationale |
|---|---|---|
| 1 | The solution must be based on globally established standards. | Standards are used to ensure the interoperability of systems in complex supply networks. |
| 2 | It must have the ability to work in a decentralised manner. | The majority of complex supply networks do not have a single entity able to manage and control them. SCTSs become decentralised and therefore any access and chain of custody verification system must also operate in a decentralised way. |
| 3 | It must prevent any leakage of sensitive information. | As noted above, traceability data is viewed as commercially sensitive and a source of competitive advantage for firms. This extends to the identity and location of the parties in transactions. Data must be kept confidential until the involved parties decide to reveal it in response to an authorised request. |
| 4 | It must prevent data misuse by malicious parties to the greatest extent possible. | Even anonymous data could be monitored by competitors or criminals to identify patterns and reveal identities. Attacks like this must be avoided. |
| 5 | It must be sector-agnostic. | Any solution must be suitable for any supply network in any sector, allowing it to become adopted as a standard. |
| 6 | It must be technology-agnostic. | Equally, any solution must be suitable for a wide range of technologies. |
| 7 | It must share traceability data and allow queries for the purposes of:<br>   a.  Chain of custody confirmation.<br>   b.  Querying of/by possibly unknown parties (intermediaries in the supply chain, NGOs, regulators and others).<br>   c.  Full transparency of data available for authorized parties | Traceability data supports critical business processes. By confirming chain of custody anonymously, the authenticity of products is guaranteed.<br>Second, by allowing unknown parties to ask permission to see data, access by nth tier suppliers or customers and regulators is supported.<br>Third, when an unknown party can prove their right to access data, full transparency should be possible |

**Solution Architecture**

In this section, we provide an overview of the solution approach and architecture. To fulfil functional requirement 1, it is assumed that supply chain traceability data is recorded as 'event data' following the EPCIS standard (GS1, 2016), however we note that a similar system can be built on other standards.

The proposed solution consists of three functional blocks:

1. Sanitisation and Hashing of traceability event data, to include hashing of data and upload to a networked (federated) platform
2. Chain of custody verification
3. Decentralised data access queries to, and authorization by, anonymous parties

The architecture consists of two main components (see Figure 1): firstly, the networked discovery service itself, which receives the sanitised and hashed events, and then enables chain verification. Second, the data transfer mechanism, which facilitates an anonymous access rights verification and subsequent data exchange between a querying and queried party.

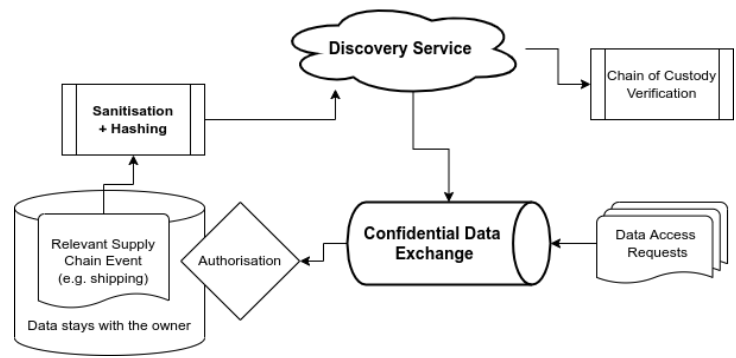

*Figure 1: Discovery Solution Architecture*

*Sanitisation of Data*

Once traceability event data is generated by an actor in the supply chain, all possibly sensitive data must be hidden. In the discovery service, this is a two-stage process: first sanitisation, then hashing. Sanitisation describes a process of systematically reducing traceability event data down to its core attributes needed for discovery and removing extraneous details such as data from sensor readings. This process also reduces the data storage requirements for the discovery service, but, more importantly, enhances privacy by data minimisation ("Datensparsamkeit") in view of requirement 3 and 4.

*Hashing*

Following sanitisation, a hashing algorithm ensures that functional requirements 3 and 4 are fulfilled also with respect to the sanitised data in a three-stage process.

First, the attributes which are to be included in the sanitised events are sorted in three categories. The values of attributes that can be shared without revealing business sensitive information (e.g. the event time and the business step) need no further treatment. The remaining values (in particular item/party identifiers) need to be hashed in order to conceal the information about who traded what with whom. Values, such as serialised (item) IDs, that contain a lot of entropy (i.e. which are very hard to guess) are hashed with a standard hashing algorithm and the hashed values (using e.g. the format as specified in RFC6920, see Farrel et al. 2013) are included in the sanitised event.

The remaining values are the most problematic ones, such as identifiers that are business relevant and need to be hidden, but at the same time are static over time. A spying competitor may "un-hash" these values by computing all hashes of all possible business IDs and comparing the outcome to the hashed data. This is a standard problem with a standard solution called "salting" where the value in question is concatenated with a high entropy string before hashing. In our case, to maintain chain of custody validation, matching events (e.g. shipping and receiving) need to be sanitised using the same salt such that the sanitised matching events contain the same hash value. This is necessary for linking the chain. We propose that an accidentally 'shared secret', such as the invoice reference, that is privately known to both parties is used as the salt.

Once sanitised and hashed, the traceability event data can be uploaded to a discovery service. In line with requirement 6, we do not stipulate the technology to be used for sharing hashed events. A DLT based solution may be used, but more conventional systems, such as a federating resolver, e.g. a GS1 Digital Link (GS1, 2022) resolver, or even a central repository can be used. No sensitive data flows through the sanitisation servers (zero trust).

*Chain of Custody Verification*

The discovery service contains all the sanitised and hashed events, shared openly, but meaningless to malicious parties due to the care taken in order to remove or conceal any sensitive information. However, if an actor has access to a product identified with a standard identification scheme such as an 'SGTIN' (GS1, 2018), the actor can hash this identifier and query the discovery service for all hashed events that contain it. All traceability events capturing a change of ownership or possession (e.g. related to a shipping or receiving process) accommodate (obfuscated) 'source' and 'destination' identifiers. By ensuring that an unbroken chain of sources and destinations exists for the product (see below), the actor is able to verify a complete chain of custody.

*Access Right Verification and Data Transfer*

Finally, the confidential data exchange service supports queries to an unknown data owner for the full traceability data (i.e. the full clear text EPCIS event). The data owner can retrieve such requests from the service and answer them if satisfied with the authorisation of the querying party. In line with our zero trust design, we propose the use of a so-called 'dead drop' approach, where an anonymous platform allows requesters to leave access requests for data owners to periodically check and respond to. Data requests contain three parts:

1. An identifier of the data that is being queried, e.g. the hashed SGTIN of the item that the data corresponds to.
2. A response protocol and endpoint, so that the requestor can receive the data. This could take many forms including an http POST + REST endpoint (URL), but also conventional means such as an e-mail address or even a phone number etc.
3. An optional set of credentials by which the querying party aims to prove its authenticity and/or authorisation to access the data.

The data owner can receive this request without revealing their identity or further traceability information and evaluate whether the requestor has access rights by regularly querying the dead drop service. If access is granted, data transfer can take place. Data transfer will happen according to the protocol asked for by the querying party in 2. In the next section, we describe an example use case for the discovery service.

**A Prototype System for Anonymous Supply Chain Traceability Data Sharing**

This section presents the software prototype developed for this paper in the format of a user story, demonstrating how two disconnected parties in the same supply chain can share sanitised data and request more information. The software prototype is available open source at (GTS 2022).

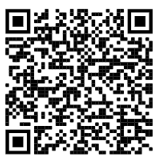 See in particular https://github.com/european-epc-competence-center/epcis-sanitisation/releases/download/v1.0.1/appendix.pdf for an extended example including full events, which are omitted from this paper's print form for brevity.

*Example Supply Chain*

The validation exercise for our software prototype draws on example data from a hypothetical electronic products supply network. Schematically represented in Figure 2, the supply network consists of six actors. In this case, a simple electronic product is assembled by *Manufacturer C* with components from *Supplier A* and *Supplier B*. The product is sold to consumers by *Retailer D* or *Retailer E*. Finally, at the end of life with the consumers, the product is returned to be recycled by *Reseller F*.

Specifically, we consider the situation where *Reseller F* receives a product and desires to verify the materials used in a component of that product. This requires *F* to be able to verify a complete chain of custody for the product, establish the existence of the suppliers, send a data access request, and receive a response from the suppliers. This provides a significant improvement over current best practice, where recyclers are able to obtain little-to-no information directly from the supply chain. Before considering the point of view of the reseller, we first demonstrate the creation of a 'commissioning' event at *Supplier A*.

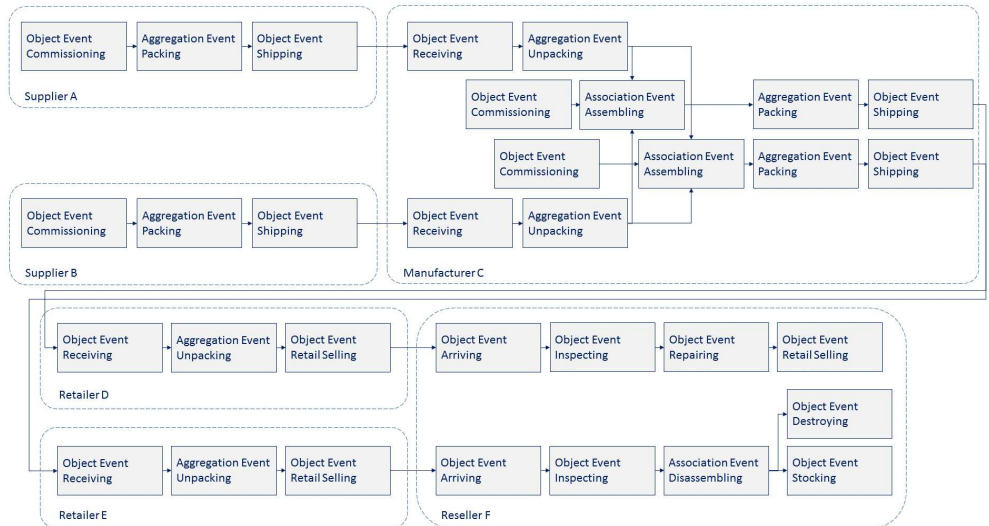

*Figure 2: Schematic of traceability events in an example supply chain. See (GTS 2022)*

### Event Creation and Sanitisation

*Supplier A* creates a key component of the electronic product (for example, the battery). This results in a commissioning event, followed by events for the packing and shipping of the battery to *Manufacturer C*. Once the traceability data is generated in the EPCIS format, *Supplier A* automatically uploads the event to an anonymous traceability data sharing system via a sanitiser. Figure 3 shows an original 'departing' event (left-hand side), its corresponding sanitised (top right) and hashed (bottom right) equivalent after being processed by our software prototype (more detail is available in the online supplement, Figure A1). The sanitised event is now public, with anyone able to see it but unable to extract sensitive information.

*Figure 3 - Overview of sanitisation. See (online) appendix for details.*

*Event Search*

The components are collated at the manufacturer, where the product is assembled and dispatched to the retailers. The retailers then sell the products to consumers. Throughout the supply network, further EPCIS traceability events are created and uploaded to the central service, but the sensitive and detailed data behind the traceability events remains decentralised and non-replicated, stored in the data owners' databases. Together, these events form a record of an unbroken chain of custody. At the end of the product's life with the consumer, it is sent to *Reseller F* who, amongst other things, recycles the battery.

To check the provenance of the product, the reseller can use the sanitised traceability events to confirm the chain of custody. First, the reseller takes the known EPC identifier of the product and hashes it with the standard hashing algorithm, which would be agreed at industry level (in this example, we use SHA-256). Then the reseller searches the traceability discovery service for this hashed EPC. The search returns all the events back to the 'association event' where the product was assembled. As this event shows the hashed EPCs of the component parts from the suppliers, the reseller can then search the traceability system for these component parts and receive all the events back to the original commissioning events.

With the series of sanitised events, it is possible for the reseller to confirm the chain of custody by connecting shipping and receiving events. This can be done anonymously in the form of a 'chain navigation algorithm', without identifying the immediate parties. From this chain of events, *Reseller F* can submit a request for more information about the commission event. See Figures A2, A3, A4 and A5 in the online supplement for code snippets showing this process for a pair of shipping and receiving events.

*Data Access Requests*

The sanitised event contains a URL where requests for data can be posted. We do not stipulate any requirements for this interface, it may be an anonymous site provided by a third party or an identifiable corporate URL. In our example case (see GTS 2022 for an implementation), the webpage allows requests to be posted publicly and searched by (hashed) EPC. It is assumed that the owners of the event data, such as *Supplier A*, continuously monitors the data requests and responds as required.

Figure 4 shows an example of a request posted by *Reseller F* asking for traceability data about the electronic product's components from *Supplier A*. The example format used to test the prototype allows a requestor to prove their right to access the data using asymmetric encryption. The data owner, in this case *Supplier A*, can choose whether to respond to the request by contacting the requestor at their specified *endpoint*. This completes the traceability data sharing process. If the data owner does not accept the request, then it is ignored, protecting their identity (in this case, it is deleted from the dead drop once it has expired

```
{
    "requesting": "dd85a8a245177fe4c4cbd540075a96dc38aefd7780677989be9e1efc92b5f08f",
    "recipient": {
        "endpoint": "https://eecc.de:234567",
        "protocol": "POST"
    },
    "auth": {
        "id": "Sebastian"
    },
    "valid_until": "2021-07-30 13:32:44"
}
```

Figure 4 – An example data access request

*Discussion of Results*

Above, we have demonstrated the prototype solution through a user story in a simple electronic products supply network. The prototype fulfils the functional requirements for anonymous supply chain traceability data sharing. It is based on globally established standards (OpenAPI, SHA-256, GTINs, & the EPCIS data model), it is decentralised, and the sanitisation process ensures no sensitive information is leaked. It is impossible for a malicious party to access any data beyond the public data in the sanitised event which can neither be linked to a product nor to any parties and which does not contain sensitive information. Further evidence of this can be seen in the live prototype, which the reader is encouraged to test (GTS 2022). Although we have tested the prototype with an electronic product example, the solution is sector-agnostic and can be hosted on many different technologies, including a distributed ledger if desired. Finally, we have shown that data can be shared and queried to support chain of custody verification and access by third parties.

Our proposed solution to the discovery problem has two limitations. First, it requires standardised adoption throughout the supply chain to be useful. To promote this, it may be required to incorporate the sanitisation approach into future industry standards for supply chain traceability. Second, the sanitisation adds another layer of complexity to traceability systems, requiring that data is processed through a sanitisation algorithm after creation. This may present a barrier to adoption.

**Conclusion**

This research paper aimed to develop a prototype solution for the problem of anonymous sharing of traceability data in a complex supply network. It was noted that current SCTSs do not support the decentralised, anonymous, and hashed sharing of traceability data. The prototype *discovery service* offers practical implications to the fields of supply chain management and information systems.

This research paper provides a complete end-to-end system for anonymously sharing and querying traceability data, verifying access rights to the data, and confirming complete chains of custody for products. The proposed solution reduces the data storage requirements to a minimum and protects data ownership rights by allowing the majority of traceability data to be stored in a non-replicated manner on the data owners' databases. Systematic hashing, including the use of salted hashes, prevent sensitive data leaking, even when large amounts of traceability data are generated. The authors note the importance of the solution being based on global supply chain information standards and being technology and industry agnostic. This maximises the commercial utility of the system, as supply networks differ significantly industry-to-industry.

Future extensions of the proposed system may include the integration of decentralised identifiers (DIDs)and digital signatures in the form of Verifiable Credentials (VCs) in order to provide court-proof evidence about the chain of custody. Blockchain technology is claimed to be valuable to SCTSs, therefore, another extension could trial hosting the discovery service on a blockchain platform, also improving the authenticity guarantees of the system.

State-of-the-art decentralised SCTSs do not have a standardised, systematic, and efficient method for anonymising data and verifying access rights. Through a design science methodology, this paper described the development of such a system, based on widely accepted global standards. Researchers should incorporate this system into future decentralised supply chain traceability system prototypes and practitioners can implement our findings to make their SCTSs both more open and more secure, simultaneously.

# A solution approach for the anonymous sharing of sensitive supply chain traceability data - Appendix

Rob Glew, Ralph Tröger, Sebastian Schmitter

# Sanitisation and Hashing of Example Events

In the sanitisation process, all unnecessary data is removed from the event and then any identifiable or possibly sensitive information is obfuscated by hashing. Hashed data is formatted according to RFC6920, i.e. a prefix like "ni://sha-256;" indicates the hashing algorithm (here sha-256). The sanitised and hashed event data is then published through the discovery service, with anyone able to see it but unable to extract sensitive information.

## Graphical Example

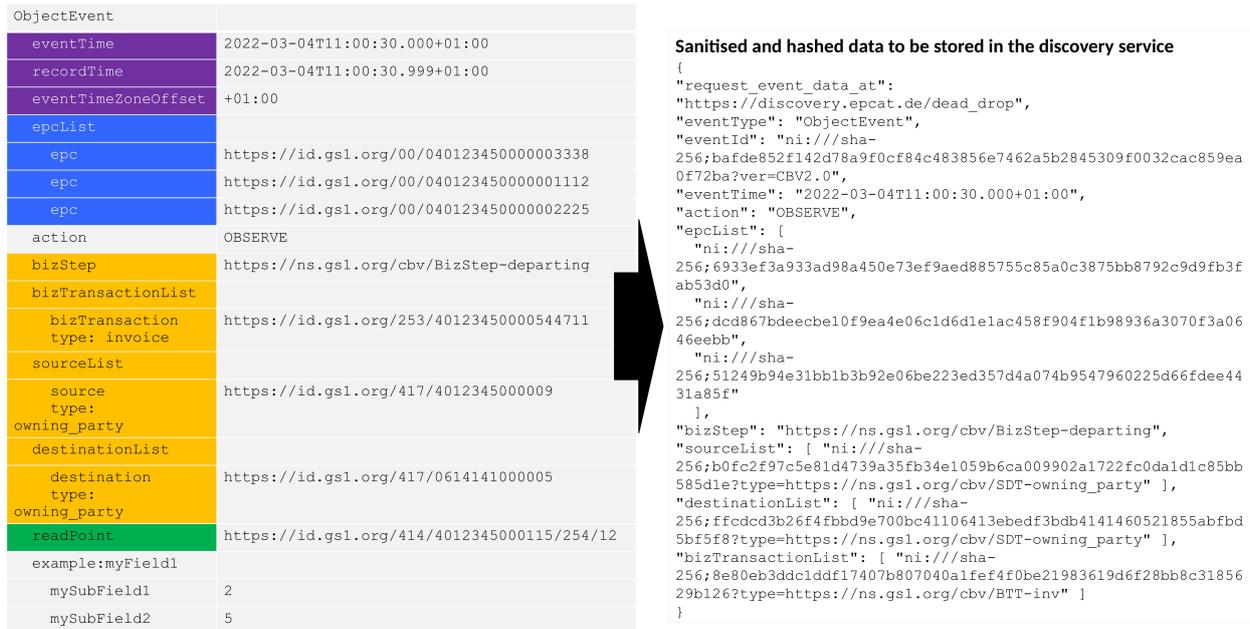

*Figure 1. Graphical Example for an EPCIS event being sanitised and hashed*

Figure 1 is a graphical representatoin of an example EPCIS event available at https://github.com/european-epc-competence-center/epcis-sanitisation/blob/master/tests/events/ReferenceEventHashAlgorithm.xml and the sanitised and hashed discovery data for this event, avaiable at https://github.com/european-epc-competence-center/epcis-sanitisation/blob/master/tests/events/ReferenceEventHashAlgorithm.sanitised.json .

# Chain Navigation

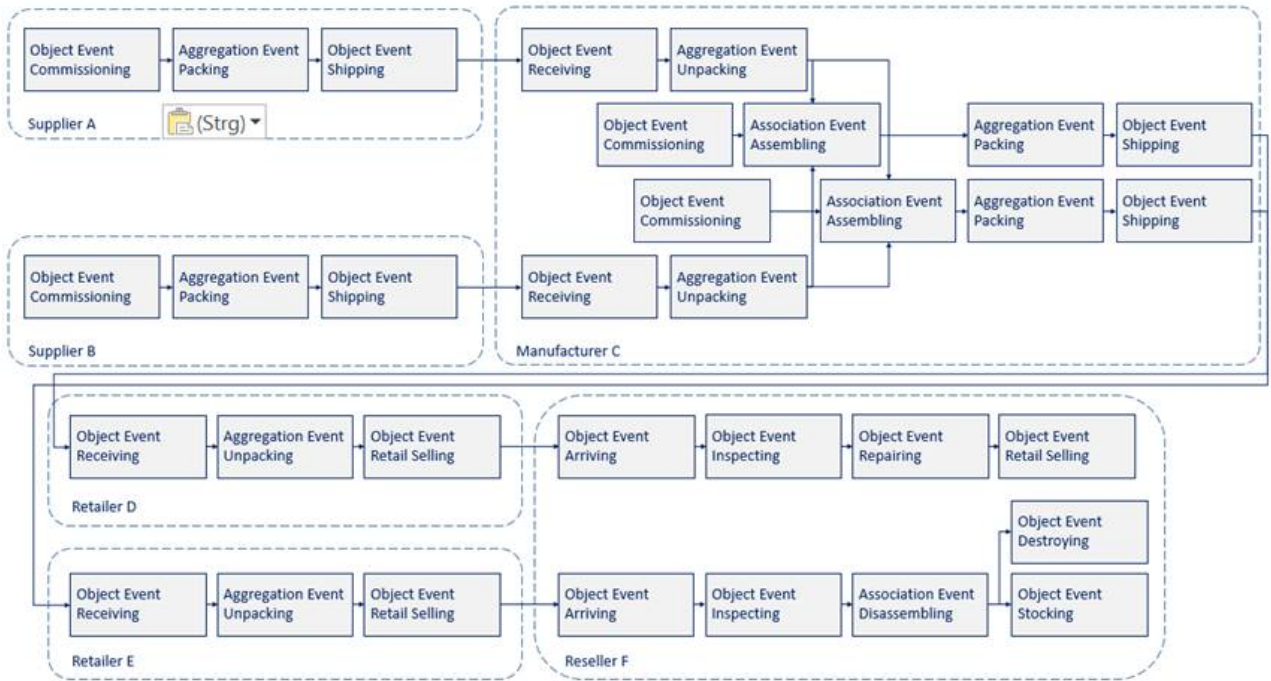

*Figure 2. Depiction of a supply chain segment*

The EPCIS events for the ficticious but realistic supply chin segment shown in Figure 2 are avaiable in full at https://github.com/european-epc-competence-center/epcis-sanitisation/blob/master/tests/events/SanitisationEventDataset.xml with the discovery data at https://github.com/european-epc-competence-center/epcis-sanitisation/blob/master/tests/events/SanitisationEventDataset.sanitised.json

We will focus on one step in this chain, naemyl the first shipping and receiving event in order to exemplyfy the chain navigation. The code snippets shown below are taken from the above mentioned larger files.

## Shipping Event

Code Block 1 shows a full EPCIS shipping event and Code Block 2 the corresponding sanitised and hashed data.

*Code Block 1. Example for an EPCIS shipping event*

```xml
<ObjectEvent>
    <eventTime>2021-04-28T00:00:00.000+02:00</eventTime>
    <eventTimeZoneOffset>+02:00</eventTimeZoneOffset>
    <epcList>
        <epc>urn:epc:id:sscc:4023333.0222222222</epc>
    </epcList>
    <action>OBSERVE</action>
    <bizStep>urn:epcglobal:cbv:bizstep:shipping</bizStep>
    <disposition>urn:epcglobal:cbv:disp:in_transit</disposition>
    <readPoint>
        <id>urn:epc:id:sgln:4023333.00002.0</id>
    </readPoint>
    <bizTransactionList>
        <bizTransaction type="urn:epcglobal:cbv:btt:po">
            urn:epc:id:gdti:0614141.00002.PO-123
        </bizTransaction>
    </bizTransactionList>
    <extension>
        <sourceList>
            <source type="urn:epcglobal:cbv:sdt:possessing_party">
                urn:epc:id:pgln:4023333.00000
            </source>
        </sourceList>
        <destinationList>
            <destination type="urn:epcglobal:cbv:sdt:possessing_party">
                urn:epc:id:pgln:0614141.00000
            </destination>
        </destinationList>
    </extension>
</ObjectEvent>
```

*Code Block 2. The santised and hashed event from Code Block 1. Also exemplifies the 'request data at' deap drop solution*

```json
{
    "request_event_data_at": "https://discovery.epcat.de/dead_drop",
    "eventType": "ObjectEvent",
    "eventId": "ni:///sha-256;8cc07dab5f2b6674fd3d892a36c9795846cad9d169bbc2d48a50cd0156c2ec41?ver=CBV2.0",
    "eventTime": "2021-04-28T00:00:00.000+02:00",
    "action": "OBSERVE",
    "epcList": [
        "ni:///sha-256;e5284a01b67b7756c0f51d10e7c74c6f277fea0e1f08ebe8f27fae25b04e695b"
    ],
    "bizStep": "urn:epcglobal:cbv:bizstep:shipping",
    "sourceList": [
        "ni:///sha-256;63ba4ead93f79fb67e68a277e85247988fb410ac0c2f00b87f802d75031b52f9?type=urn:epcglobal:cbv:sdt:possessing_party"
    ],
    "destinationList": [
        "ni:///sha-256;8d2cdc63d2e3d173174c9167ac4a857dfc0a0abba7cee54ef0e4b9a21156021b?type=urn:epcglobal:cbv:sdt:possessing_party"
    ],
    "bizTransactionList": [
        "ni:///sha-256;2428dd1fddb2811d950320b732dda8f4be7312e02be14c2dfb8da9969085da38?type=urn:epcglobal:cbv:btt:po"
    ]
}
```

## Receiving Event

Code Block 3 shows a full EPCIS shipping event and Code Block 4 the corresponding sanitised and hashed data.

*Code Block 3. Example for an EPCIS receiving event*

```
<ObjectEvent>
    <eventTime>2021-04-29T00:00:00.000+02:00</eventTime>
    <eventTimeZoneOffset>+02:00</eventTimeZoneOffset>
    <epcList>
        <epc>urn:epc:id:sscc:4023333.0222222222</epc>
    </epcList>
    <action>OBSERVE</action>
    <bizStep>urn:epcglobal:cbv:bizstep:receiving</bizStep>
    <disposition>urn:epcglobal:cbv:disp:in_progress</disposition>
    <readPoint>
        <id>urn:epc:id:sgln:0614141.00012.0</id>
    </readPoint>
    <bizTransactionList>
        <bizTransaction type="urn:epcglobal:cbv:btt:po">
            urn:epc:id:gdti:0614141.00002.PO-123
        </bizTransaction>
    </bizTransactionList>
    <extension>
        <sourceList>
            <source type="urn:epcglobal:cbv:sdt:possessing_party">
                urn:epc:id:pgln:4023333.00000
            </source>
        </sourceList>
        <destinationList>
            <destination type="urn:epcglobal:cbv:sdt:possessing_party">
                urn:epc:id:pgln:0614141.00000
            </destination>
        </destinationList>
    </extension>
</ObjectEvent>
```

*Code Block 4. The santised and hashed event from Code Block 3*

```json
{
    "request_event_data_at": "https://discovery.epcat.de/dead_drop",
    "eventType": "ObjectEvent",
    "eventId": "ni:///sha-
256;7a742a2be1d9c5cc71bda8d58abc93393236e54deca8b5d1672acc8022d6ec34?ver=CBV2.0",
    "eventTime": "2021-04-29T00:00:00.000+02:00",
    "action": "OBSERVE",
    "epcList": [
        "ni:///sha-
256;e5284a01b67b7756c0f51d10e7c74c6f277fea0e1f08ebe8f27fae25b04e695b"
    ],
    "bizStep": "urn:epcglobal:cbv:bizstep:receiving",
    "sourceList": [
        "ni:///sha-
256;63ba4ead93f79fb67e68a277e85247988fb410ac0c2f00b87f802d75031b52f9?type=urn:epcgloba
l:cbv:sdt:possessing_party"
    ],
    "destinationList": [
        "ni:///sha-
256;8d2cdc63d2e3d173174c9167ac4a857dfc0a0abba7cee54ef0e4b9a21156021b?type=urn:epcgloba
l:cbv:sdt:possessing_party"
    ],
    "bizTransactionList": [
        "ni:///sha-
256;2428dd1fddb2811d950320b732dda8f4be7312e02be14c2dfb8da9969085da38?type=urn:epcgloba
l:cbv:btt:po"
    ]
}
```

## Linking the Chain

The sanitised and hashed data in the above examples Code Block 2 and Code Block 4 is shared through the Discovery Service. Anyone who wants to obtain information about the shipment with the id `urn:epc:id:sscc:4023333.0222222222` (see Code Block 1) can now hash the id with standard tools like `echo -n "urn:epc:id:sscc:4023333.0222222222"|sha256sum` to obtain the hex encoded sha256 hash `e5284a01b67b7756c0f51d10e7c74c6f277fea0e1f08ebe8f27fae25b04e695b` and query the discovery service to obtain the data sets Code Block 2 and Code Block 4. From these dataset it can be verified that there are matching shipping and receiving IDs of parties with type `posessing_party`. This enables to establish an anonymous chain of custody.

## Zero Trust Clear Text Exchange

Using the `request_event_data_at` url pointing to the confidential data echange service, anyone can post a request for more information. The data owner may then regularly query for such requests, decide on their legitimacy, and provide all or some clear text information from the original event to the querying party if authorised.

If the data owner finds a request to be insufficiently authorized, he can just ignore the request, which will then time out. This ensures that no information what so ever about the data owner or any of the parties involved in the business transactions is revealed.